# Impacts of Illuminance and Correlated Color Temperature on Cognitive Performance: A VR-Lighting Study

**Armin Mostafavi [1,2], Milica Vujovic [1], Tong Bill Xu [2], Michael Hensel [1]**

[1] Research Department for Digital Architecture and Planning, Vienna University of Technology, Vienna, Austria.

[2] Department of Human Centered Design, Cornell University, Ithaca, NY, USA.

**Abbreviations:**

AC Architectural Context

BDST Backward Digit Span Task

CB Cold-Bright lighting condition

CD Cold-Dark lighting condition

DST Digit Span Test

ET Experiment Time

HMD Head-Mounted Display

LC Lighting Condition

NM Neutral-Moderate lighting condition

RQ Research Question

TMO Tone Mapping Operator

RT Reaction (response) Time

SAM Self-Assessment Manikin

UE4 Unreal Engine 4

VMT Visual Memory Task

VE Virtual Environment

VR Virtual Reality

WB Warm-Bright lighting condition

WD Warm-Dark lighting condition

**Abstract:**

This study contributes to the ongoing exploration of methods to enhance the environmental design, cognitive function, and overall wellbeing, primarily focusing on understanding the modulation of human cognitive performance by artificial lighting conditions. In this investigation, participants (N=35) engaged with two distinct architectural contexts, each featuring five different lighting conditions within a virtual environment during specific daytime scenarios. Responding to a series of cognitive memory tests, we measured participant test scores and the corresponding reaction time. The study's findings, particularly in Backward Digit Span Tasks (BDST) and Visual Memory Tasks (VMT), indicate that diverse lighting conditions significantly impacted cognitive performance at different times of the day. Notably, the BDST scores were mainly affected by lighting conditions in the afternoon session, whereas the VMT scores were primarily influenced in the morning sessions. This research offers support for architects and engineers as they develop lighting designs that are sensitive to the cognitive performance of occupants. It highlights the advantages of utilizing VR simulations in the AEC industry to assess the impact of lighting designs on users. Further research can lead to the development of lighting systems that can promote better cognitive function and overall wellbeing.

## 1. Introduction

Light plays a crucial role in human health and cognitive performance. The health effects of poor artificial lighting in indoor environments can be profound [1]. Eye strain, headaches, exhaustion, and mood disorders are just a few issues that can arise from poor lighting design [2], [3], [4], [5]. Illuminance and Correlated Color Temperature (CCT) are essential lighting design and evaluation metrics. Light intensity and color temperature can significantly impact a person's physiological and psychological wellbeing [6], [7]. Illuminance is a measure of the amount of light that falls on a surface, and it is measured in lux [8], while CCT is a metric that indicates the color appearance of light [9].

Studies have shown that exposure to bright light during the day can help regulate the human circadian rhythm, improving sleep patterns and overall health [10], [11]. Conversely, exposure to blue light, which has a higher CCT, in the evening can disrupt the body's natural sleep cycle and lead to sleep disturbances [10]. In addition, insufficient exposure to light during the day can lead to mood disorders, such as depression and seasonal affective disorder [10], [12]. The CCT of light can also impact cognitive performance [13], [14], [15], [16]. For example, exposure to cooler, bluish-white light can improve alertness and concentration, whereas exposure to warmer, yellowish-white light can induce relaxation and reduce stress. In addition to light intensity and CCT, the timing and duration of exposure can also impact health and cognitive performance [17], [18]. Prior research has shown varied outcomes regarding the impact of CCT and illuminance on cognitive performance, with some studies indicating positive effects ([13], [18], [19], [20], [21], [22]), others finding null or no significant impact ([13], [18], [23], [24], [25], [26]), and a few reporting negative effects ([24], [27], [28]).

This study underscores the pivotal role of lighting in both human health and cognitive performance, emphasizing that its effects are contingent on various factors, including illuminance, CCT, daytime conditions, and reaction time [29], [30], [31]. Additionally, the introduction section highlights virtual reality (VR) integration as a valuable tool for precisely controlled lighting experiments. Recognizing and incorporating these factors is crucial when devising lighting environments, especially in settings where health and cognitive performance are of paramount importance, such as schools, hospitals, or workplaces [22], [32], [33], [34], [35].

**1.1 Lighting and Cognitive Performance**

Lighting's impact on cognition and memory is extensively studied [28], [36], [37], [38]. Interdisciplinary research that combines architecture, cognition, and psychology [39], [40] shows how bright natural light (3000lx) improves mood and cognitive performance, while poor lighting leads to issues like depression and reduced cognitive function [19]. Furthermore, studies show that bright light (900lx vs. 90lx) enhances attention, while dim light causes drowsiness [31]. Research conducted in an office environment shows that light of high/medium correlated color temperature enhances perception, learning, and memory [41]. On the other hand, studies on lighting correlated color temperature's impact on comfort and cognitive performance indicate significant effects on memory test scores, with better performance at 650 lx compared to 1050 lx [13].

Furthermore, classroom studies found that lighting significantly affects cognition, learning, and memory [13], [42]. Research in Virtual Learning Environments (VLEs) reveals that high illuminance improves students' memory, attention, and executive functions. Blue lighting in VLEs enhances short-term memory and executive functions [43]. However, another research on immersive virtual environments shows that low illuminance (100 lx) improves memory performance compared to higher illuminance levels [42].

Current investigations lack consensus on illuminance and CCT's specific impact on cognitive domains [23]. Methodological variability complicates outcomes' generalization. For instance, the comparison of lighting results obtained in virtual environments [42], [44], [45], [46] with those in physical environments [23], [30], [47] could be another barrier to having a comprehensive attitude toward the effects of LCs on user cognitive performance. It is crucial to note that lighting influences both visual and auditory systems, impacting overall cognitive functioning and audio memory [47].

**1.2 Lighting and Reaction Time**

Lighting significantly influences reaction time [47]. The visual system's dependence on light is crucial for efficient functioning, with well-lit environments leading to faster and more accurate visual processing [48], [49]. Ample light triggers rapid chemical reactions in the retina, facilitating swift responses to stimuli [50], [51], [52]. Exposure to bright light or changes in lighting conditions can affect our circadian rhythms and alter our physiological state, which can impact our reaction time [53], [54], [55]. For instance, exposure to bright light in the morning can help to improve alertness and reaction time, while exposure to bright light in the evening can disrupt our sleep-wake cycle and lead to slower reaction times the next day.

Conversely, poorly lit settings with reduced visibility and increased visual noise can slow reaction times [49], [56], as details may be overlooked, delaying visual information processing. Motor responses, including reaction time, result from the integration of visual information and cognitive processing [56]. Cognition, closely tied to visual processing, depends on light quality and availability [57]. The speed and accuracy of visual processing, influenced by lighting conditions, directly impact cognitive performance [49]. Understanding the relationship between lighting conditions and cognition can guide the design of environments for enhanced cognitive performance.

**1.3 Lighting Conditions and Daytime**

Numerous studies investigating the relationship between lighting conditions and time of day with cognitive performance reveal important dependencies, offering implications for lighting and architectural research [17], [24], [47], [58]. Huiberts et al. [17] explored the correlation between illuminance level, task difficulty, and time of day in cognitive task performance, employing auditory working memory tasks like Digit Span and n-back tasks. Bright light was found to decrease morning sleepiness but not in the afternoon. Subjects reported feeling more energetic under bright (1700lx) morning light compared to dim (165lx) light, with an opposite effect observed in the afternoon. Complex tasks (7–8 digits vs. 4-6 digits BDST) exhibited afternoon declines under bright light, while less complex tasks showed improvements. Notably, conclusive insights regarding the time of day and bright light's cognitive performance benefits remain elusive [14].

Additional research by Huiberts et al. [30] focused on lighting conditions' effects on cognitive performance using the Psychomotor Vigilance Task and backward digit-span task. Regardless of the time of day, Backwards Digit-Span Task performance significantly improved under brighter light. Though activities preceding the Backwards Digit-Span Task varied, a potential association between overall cognitive load, lighting conditions, and time of day emerged. Subsequent studies affirmed bright light's positive impact on intricate BDST tasks and reaction time on the Psychomotor Vigilance Task, indicating an overall advantageous effect.

Chen et al. [59] discovered that participants preferred 500 lx illuminance during the morning and afternoon, but lower illuminance levels (50 lx and 100 lx) were more comfortable in the evening. The study underscores the importance of considering daytime effects when assessing lighting conditions for subjective comfort, especially during daytime activities. Research by Smolders et al. [24] suggests that correlated color temperature's impact on alertness, cognitive performance, and arousal varies with the time of day. Exposure to light with a correlated color temperature of 6000 K, compared to 2700 K, led to decreased positive affect and increased negative affect regardless of exposure timing. Higher correlated color temperature levels were associated with diminished emotional wellbeing and a less favorable perception of the environment.

This study aims to assess cognitive performance in users by utilizing established working memory tests, as suggested and detailed in the literature reviewed in previous sections. Specifically, we will employ the Backward Digit Span Tasks (BDST) and Visual Memory Tasks (VMT) for this purpose. The evaluation of cognitive performance, with a particular emphasis on short-term and working memory, will hinge on two main metrics: the scores achieved in these tests and the reaction times recorded for the participants.

**1.4 Lighting and Virtual Reality**

The Architecture, Engineering, and Construction (AEC) industry heavily relies on digital modeling, virtual simulations, and visual communication tools to ensure successful and efficient project completion at a high level of quality [60]. In studies exploring lighting designs, researchers employ VR head-mounted displays (HMDs) to immerse participants in a generated 3D model, swiftly collecting user responses. This VR method allows seamless switching between various CCTs and illuminance levels, offering a controlled environment coupled with human subjective behavior and biofeedback sensors, like EEG, which is impractical in real-world settings [42], [61], [62]. In the realm of human-centered lighting, VR serves as a valuable instrument for evaluating visual quality and lighting perception [44], [63], [64]. Modern VR-HMDs provide a realistic visual experience, with a growing body of literature comparing perceptual accuracy with real-world lighting settings [42], [65], [66], [67], [68], [69], [70], [71].

In a seminal VR lighting study, Heydarian et al. [69] demonstrated that immersive virtual environments effectively measure lighting preference, comparing physical environment (PE) vs. VE. Chen et al. [68] similarly compared lighting settings in PE vs VE, concluding that VR presents lighting attributes consistently with PE, making it a viable solution for scientific investigation. Hong et al.'s study [70] investigated the adequacy of VE representation vs. PE in windowed office space, revealing a great sense of presence in VR with no discernible difference in occupant satisfaction. However, VR displays limitations in luminance range, impacting effectiveness in extremely high or low illuminance situations [72], [73].

It is crucial to note that VR displays cannot reproduce the same high intensities as real-world light sources, hindering accurate simulation of blinding effects and glare. VR falls short in depicting low-light settings accurately, as current displays cannot reproduce scotopic vision range conditions, necessitating artificial reproduction of color shifts and diminished visual acuity [70].

Recent VR lighting studies address dynamic range limitations. Abd-Alhamid et al.'s experiment [65] investigated subjective and objective visual responses, finding no significant difference in perceived lighting and room sensations but differences in visual-quality attributes assessment. Rockcastle et al.'s study [67] suggested contemporary VR HMDs could be reasonable surrogates for well-lit lighting scenes in subjective evaluations of visual comfort, pleasantness, evenness, contrast, and glare. Despite limitations, VR HMDs remain valuable for lighting research, providing an alternate representation medium for accurate scene appearance acquisition in visual perceptions [63].

**1.5 Study Goals and Research Questions**

Further investigation is needed to understand the intricate relationship between lighting conditions and cognitive function. There is a significant research gap in comprehending the impact of various light characteristics on human behavior and cognitive performance, with a broader objective of enhancing knowledge of how light influences human health [1]. Based on former studies introduced in section 1, this research hypothesizes that cognitive performance is significantly influenced by the interplay of lighting conditions (i.e., light intensity and CCT), and the effects can be moderated by "time of the day" (daytime) and "architectural context". The study aims to analyze and elucidate the complex relationships between various lighting parameters and cognitive performance (i.e., measuring the visual and auditory test scores and the corresponding response time) by conducting cognitive tests in virtual reality simulations of relevant settings. Hence, the primary objective of this study is to analyze the interdependencies between lighting conditions and cognitive performance. The following research questions (RQs) arise from the literature and the presented research aims:

**RQ1a:** How do different lighting conditions influence auditory memory "Scores"?
**RQ1b:** How do different lighting conditions influence visual memory "Scores"?
Additionally, RQ1 examines whether architectural context (RQ1c) and experiment time (RQ1d) moderate the influence of lighting conditions on memory "Scores."

Based on the former studies on the effect of light on cognitive performance in the real environment [23], [29], [58], this study investigates and expands research on the effects of light in VR for both auditory and visual tests. Moreover, the study is looking into the moderation effects of architectural context and layout of the room, which could alter the lighting perception of the environment based on former studies [42], [74], [75], [76], [77]. Additionally, former studies show that experiencing different indoor LCs during the daytime will result in different cognitive performances [30], [47].

**RQ2a:** How do different lighting conditions influence auditory memory test "Response Time"?
**RQ2b:** How do different lighting conditions influence visual memory "Response Time"?
Additionally, RQ2 examines whether architectural context (RQ2c) and experiment time (RQ2d) moderate the influence of lighting conditions on memory "Response Time."

While previous studies addressed the variation of Response Time (or response speed) associated with a cognitive task in different lighting conditions in the physical environment [23], [47], [78], this study replicates those measurements during participant interaction under different LCs in VR which allows testing the influence of different factors systematically.

It should be considered that the use of VR itself is not a novelty, and we emphasize that the unique contribution of our study is the systematic examination of how different factors within a VR environment interact and influence cognitive performance. The utilization of VR allows for the manipulation of architectural context, lighting parameters, and task difficulty in a controlled and immersive setting, providing valuable insights into their combined effects.

## 2. Methods

### 2.1 Virtual Environment Development

We designed two different contexts in a VR environment: a modern interior classroom and a meeting room with exact interior dimensions of 7m x 8m x 3m (L x W x H). We kept the luminaires (N=12), window, door, and projection area in the same position and shape while we changed the room furniture and layout to convey the function of the space as a classroom or office. Most of the work for this virtual environment, including modeling and UV mapping, was done using Autodesk 3ds Max. The Unreal Engine 4.26 was used for texturing, lighting, and VR interactions. The Blueprint platform was utilized for front-end interaction and user interactivity. Participants could explore the virtual office environment while seated at a desk and had the freedom to look around the room. The participants' position with respect to the walls and the projection area was consistent across both environments. They were positioned at a fixed distance of 5 meters from the screen and 3.5 meters from the adjacent wall.

Participants were introduced to the VR setting by utilizing a Meta Quest 2 head-mounted display with Fast Switch Liquid-crystal Displays connected to a gaming laptop. The VR environment had a resolution of 1832 x 1920 pixels per eye and a brightness capability of up to 100 Nits (equivalent to approximately 342.6 lumens). The Meta Quest 2 headset offered a 90° horizontal field of view and a refresh rate of 90 Hz. The Meta Quest 2 displays closely follow the Rec.709 color space's RGB primaries while still using a white point that is very close to D75 [79]. The VR HMD was connected via cable to the Schenker DRT laptop (GPU: NVIDIA GeForce RTX 2080Super-8GB GDDR6, CPU: Intel Core i9-10900K RAM: 128 GB DDR4-3200) using Meta Quest Link [79].

## 2.2 Lighting Conditions

The pre-processing lighting simulation was conducted using DIALux Evo software with the CoreLine Recessed profile. We designed five different lighting conditions ranging from 600 lm to 4200 lm in luminous flux and 2000 K to 7200 K in color temperature for all 12 luminaires (all artificial lights in the simulated office). The comparison of luminance values between Unreal Engine and DIALux Evo was based on the Scorpio et al. study [80], which shows that Unreal Engine can reproduce light distribution correctly with proper calibration. According to our calculation in DIALUX Evo, the result of using 600 lm, 2400 lm, and 4200 lm for the simulated lighting (12 luminaires) was equal to an average of 100 lx, 400 lx, and 700 lx on the work plane, respectively. Hence, the produced five lighting conditions were Warm-Dark (2000 K-600 lm), Warm-Bright (2000 K-4200 lm), Cold-Dark (7200 K-600 lm), Cold-Bright (7200 K-4200 lm), and Neutral-Moderate brightness (4600 K-2400 lm) (Table 1). This gave us the possibility of a two-by-two comparison for both illuminance and CCT as these five lighting conditions had 1) Warm vs. Cold conditions, 2) Bright vs. Dark conditions, and 3) Both illuminance and CCT vs. Neutral conditions. Ultimately, the same number of luminaires with a light distribution profile was imported and simulated in UE4. The right and left-hand controls were assigned to the lumen and color temperature during the experiment's adjustment phase. The maximum amount of luminance from the VR-HMD right lens is measured using a Minolta LS-110 luminance meter, and the values are reported in section 6.1.

The available illumination level (lux) at a surface like a desk cannot be accurately read using Unreal Engine 4.26. (UE4) As a result, we chose to measure brightness using the strength of the lighting source (lumens) [81]. Choosing the lumen as the main light unit in UE4 was also explained and compared in the Scorpio et al. [80] study in a subsection known as method three. We opted for UE4 over other engines due to several reasons. Firstly, UE4 incorporates lighting algorithms based on physically based shading, ensuring accurate simulation of the interaction between light rays and surfaces following the inverse square law [82], including interactions between light and materials [83], [84]. Secondly, UE4 employs a set of physically based lighting units [81]. Additionally, there is no requirement for extra simulation software or HDR cameras to assess or replicate the distribution of brightness on the inner surfaces of the virtual setting.

Consequently, individuals can enter the illuminated virtual environment and assess the lighting design from various perspectives without any additional equipment. This enables a more comprehensive evaluation of user satisfaction, comfort, and interaction within the virtual constructed environment. It should be noted that the simulation of lighting in VR engines is a challenging programming task and is evolving rapidly [85]. Unreal Engine 4.26 employs pre-computed lighting and post-processing algorithms to improve visual quality. We followed the guidelines suggested in the engine's documentation by extending the default luminance range to set our brightness conditions, which allowed us to use correct lux values for lights in the scene and have them respected by auto exposure without causing the image to be blown out [81], [86], [87].

A Tone Mapping Operator (TMO) was employed to convert High Dynamic Range (HDR) scenes compatible with VR HMD. Our TMO for Low Dynamic Range (LDR) is similar to the one used in Hegazy et al.'s experiment [87] Although the exploration of tone mapping techniques in lighting research using Interactive Virtual Environments (IVEs) is still ongoing [66], [67], [88], [89], [90], [91], we used the Academy Color Encoding System [92] (ACES) Filmic Tone Mapping Curve, currently a default tone mapping curve in UE4 [93], [94]. The ACES Filmic tone mapping algorithms ensure the preservation of consistent color across different formats and display devices. This approach maintains color accuracy and serves as a measure to future-proof the source material. As a result, there is no need for constant adjustments for each new medium that emerges, saving time and effort [95]. In the developed system, the Filmic TMO was defined by the following parameters: Slope = 0.88, Toe = 0.55, Shoulder = 0.26, Black clip = 0.0, White clip = 0.04. Gamma correction (2.2) was implemented as part of the rendering pipeline in UE4 [79]. While we made efforts to create realistic lighting simulations using UE4, it is important to acknowledge that further research, such as this study, is required to validate and discuss these findings in relation to real-world environments.

**Table 1.** Lighting condition specifications for twelve luminaires in VE.

| Lighting Condition | Luminance (lm) | Average Illuminance (lx) | Correlated Color Temperature (K) |
|---|---|---|---|
| **Warm-Dark (WD)** | 600 | 100 | 2000 |
| **Cold-Dark (CD)** | 600 | 100 | 7200 |
| **Neutral-Moderate (NM)** | 2400 | 400 | 4600 |
| **Warm-Bright (WB)** | 4200 | 700 | 2000 |
| **Cold-Bright (CB)** | 4200 | 700 | 7200 |

**Figure 1.** Experiment LC from the participant's view, contexts, and a participant during data collection. **A.** Warm and Dark (WD) condition. **B.** Cold and Dark condition (CD). **C.** Neutral and Moderate brightness condition (NM). **D.** Warm and Bright condition (WB). **E.** Cold and Bright condition (CB). **F.** A participant during data collection. **G**. Classroom environment diagram **H.** Meeting room environment diagram.

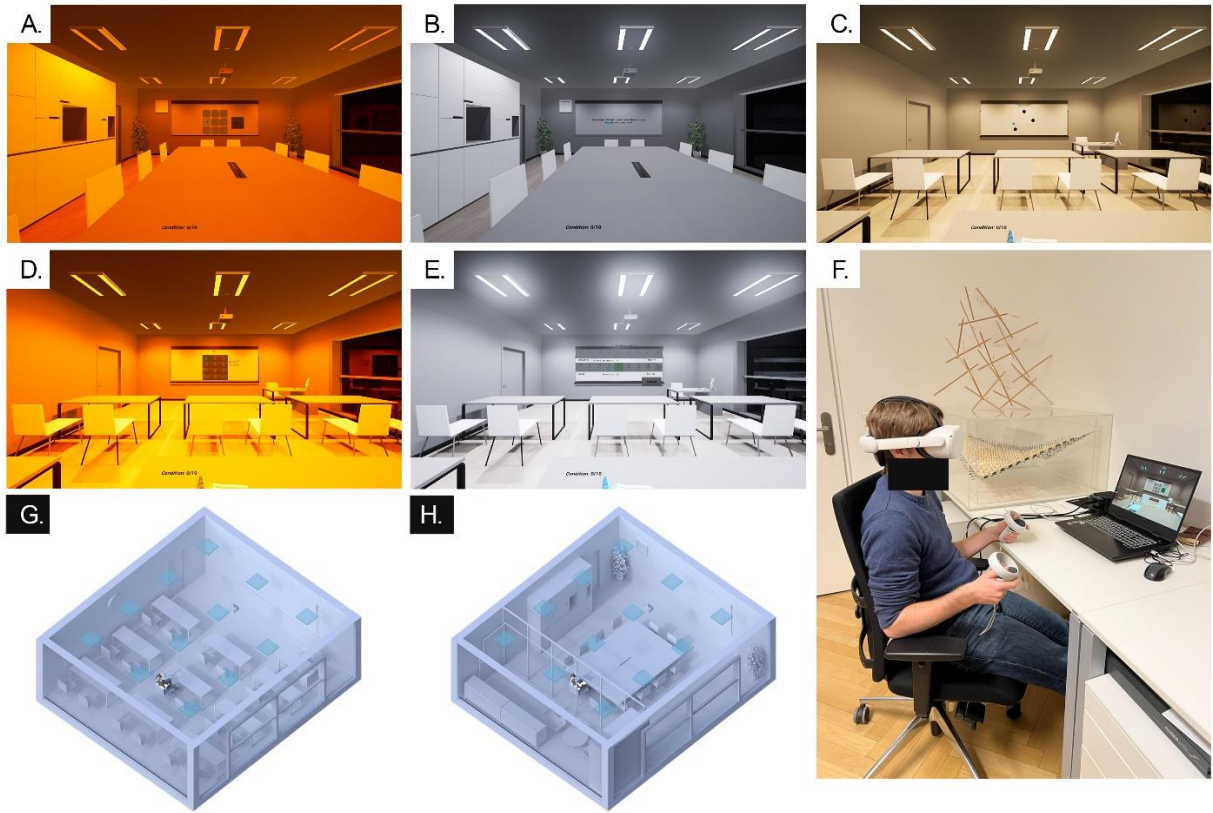

## 2.3 Participants

Thirty-five healthy adult participants were recruited using a convenience sampling method (word-of-mouth and announcements through the university transmission system). Most of the participants were associated with the **Vienna University of Technology.** The participants ranged in age from 19 to 38 years (M=26.51, SD=4.31). The majority were students and researchers (n=24), with a smaller number of engineers (n=4) and programmers (n=3). In terms of their gender, 19 participants reported as male, 15 reported as female, and one as non-binary. Each participant gave informed written consent prior to the experiment, and the overall protocol was approved by the Pilot Research Ethics Committee of **[Deleted for the purpose of blind review]**. Around half of the participants (48.5%) reported using eyeglasses or contact lenses. All participants affirmed that they had sufficient sleep (more than 7 hours) and no medical problems, such as cataracts, glaucoma, injury, or difficulty distinguishing colors.

All the experiment sessions took place in the same physical location at the Department of Digital Architecture and Planning. Sessions were conducted for one participant at a time. We conducted the experiment in the 90-minute timeslots from 9:00 am to 18:00 pm. For the data analysis, the participants were allocated randomly to three distinct timeslot groups known as **morning** (9:00 to 12:00, N=12), **afternoon** (12:00 to 15:00, N=13), and **evening** (15:00 to 18:00, N=10) to investigate the effect of time of the day on the scores and RTs. All data were collected during the first two weeks of December 2022. There was no significant difference in age and gender between daytime groups (table 2).

**Table 2.** Demographic Characteristics by Daytime.

|  | Morning | Afternoon | Evening | Overall | Difference |
|---|---|---|---|---|---|
| **Age** | 25.92 (3.38) | 26.15 (4.93) | 27.89 (4.76) | 26.51 (4.31) | $F(2, 32) = 0.61, p = .549$ |
| **Gender** | | | | | |
| Man | 7 (54 %) | 7 (54 %) | 5 (56 %) | 19 (54 %) | $p = .658$ (Fisher's Exact Test) |
| Non-binary | 0 (0 %) | 0 (0 %) | 1 (11 %) | 1 (3 %) | |
| Woman | 6 (46 %) | 6 (46 %) | 3 (33 %) | 15 (43 %) | |

Note: Age: Mean (SD). Gender: N (%).

## 2.4. Measurement Tools and Metrics

We used a variety of measurement tools from previous research, some of which were in the form of questionnaires and researchers' observations. Regarding the auditory memory test, we used Backward Digit Span Task (BDST) [96], [97] (ranging between 0-1), and for visual memory, we used Visual Dot Task (VDT) [98], [99], [100] to assess STM (ranging between 0-1) and rank the cognitive performance. We calculated the BDST score based on the sum of three different scales: a) We considered the scores as 1 (multiplied by the number of digits) for fully correct answers and 0 for any other answer based on the method used by Unsworth and Engle [101], b) correct in any-order, the sum of digits correctly recalled regardless of their serial position, and c) correct in serial-order, the total number of digits correctly recalled in the exact serial position adopted from Wambach et al. [102]. Therefore, the final score is calculated between 0 (not scored at all) and 1 (correct answer) by calculating the average of points a, b, and c. The scoring logic for VMT was based on the Fletcher experiment [103], calculated as the sum of the correct dot positions recalled by participants. For both BDST and VMT, we measured the reaction time as a second metric to study cognitive performance [102], [104], [105]. We measured the response time in milliseconds, which is the time users spent answering each question (ranging between 0 – 15 seconds).

We also used the Morningness–Eveningness Questionnaire (MEQ) [106] to assess participants' chronotype and their alertness peaks (score range 16-86). Igroup Presence Questionnaire (IPQ) [107] was measured (range 1-7) for the sense of presence experienced in a virtual environment, along with the Simulator Sickness Questionnaire [108] for assessing the users' level of cybersickness symptoms after experiencing the virtual setting. Additionally, The NASA Task Load Index questionnaire [109] was conducted to assess the subjective mental workload of the participant at the end of the experiment (1-10 scale). Our measurements of emotional valence and arousal used the Self-Assessment Manikin (SAM) [110], which is a Likert-type scale ranging from 1-9. Participants were instructed to rate their emotional valence as "how unhappy (1) or happy (9) you feel about performing in the given lighting condition and given context," while being unhappy will be interpreted as a negative and happy as a positive effect of given light for the context. Regarding arousal, they were asked to rate "how calm (1) or excited (9) do you feel about the given lighting condition and given context," while being calm or excited will neither be interpreted as negative nor positive.

Lastly, the level of immersion after the VR session was measured using the Presence Measurement, Effects, Conditions of Spatial Presence Questionnaire (MEC-SPC) [111]. This instrument is a multidimensional measure of spatial presence (ranging from 1-5) and its components, which in our case include three dimensions, each with a 4-item scale including Spatial Situation Model (SSM), Spatial Presence- Self Location (SPSL), and Sapatial Presense-Possible Actions (SPPA) with the response ranging from 1 (strongly disagree) – 5 (strongly agree).

**2.5 Power Analysis and Sample Size**

The unit of analysis was trials/tasks nested within participants. Assuming an intra-cluster correlation (ICC) of 0.10, given 20 trials per participant, the design effect was estimated to be 1+(20-1)*0.10=2.9. With 35 participants collected, the effective sample size was estimated to be 35*20/2.9=241.38. Assuming medium effect size (f=.25) and at an alpha level of .05, for effects of five different lighting conditions on dependent measures (RQ1ab, RQ2ab), f-tests would achieve a power of .88; for the lighting condition by architectural context interaction (RQ1c,

RQ2c), the achieved power was estimated to be .88; for the lighting condition by experimental time (RQ1d, RQ2d) interaction effects, we would achieve a power of 0.78.

## 2.6 Procedure

This research is focused on two educational environments - a traditional classroom and a meeting room- that are simulated in VR. Moreover, we introduced five different lighting conditions in each of the simulated environments and thus obtained ten setups in which we tested the participants' cognitive abilities (within-subjects). Moreover, we randomly assigned participants to different daytime timeslots to assess the moderation effect of experiment time (between-subjects) while controlling for the order effect of light and task difficulty. The use of within/between-subjects evaluation aligns with methodologies employed in similar lighting cognitive studies conducted in laboratory settings (see [17], [30], [44]). It should be noted that the two task difficulty levels in this experiment are relatively easy (see section 3). The examination of cognitive abilities consisted of well-known tests that engage working memory. The experimental setup was established in order to understand correlations between variables and provide answers to the research questions.

The experiment had three parts: 1) signing the consent form and pre-questionnaire, 2) the VR session, and 3) the post-questionnaire with the exit interview (Fig. 2). In part 1, after providing consent, participants were asked to complete a pre-experiment survey. Then, for part 2, the researchers equipped the participant with a VR headset. At the beginning of the VR session, all participants were instructed about the VR experiment sequence, emotional valence and arousal interpretation, memory tests, and adjustment tasks. Participants were then sequentially exposed to ten different lighting conditions, with the order of conditions randomized for each participant. At the beginning of each VR session (step 0), the participant was given a ~3-min tutorial period to become comfortable with the experiment sequence, widget interaction, and lighting adjustment tools, as well as to ensure the proper sound level and visual quality check by reading the introduction widget texts on the front board.

Each participant first experienced each condition without any task included and was free to look around and analyze the given LC (step 1). After ten seconds, we asked participants to rate their emotions using the SAM 9 scale Likert (step 2). Participants were instructed to rate their emotional valence as "how unhappy (1) or happy (9) you feel about performing in the given lighting condition and given context," while being unhappy will be interpreted as a negative and happy as a positive effect of given light for the context. Regarding arousal, they were asked to rate "how calm (1) or excited (9) do you feel about the given lighting condition and given context," while being calm or excited will neither be interpreted as negative nor positive. It should be noted that further explanation was given (in step 0) to the participants for the difference between arousal and valence based on Russell's Circumplex model for Valence-Arousal (VA) [115]. The researcher explained to the participant that arousal (intensity) reflects the degree of autonomic activation an event (i.e., given lighting condition) triggers, spanning from calm or low to excited or high. Conversely, valence denotes the extent of pleasantness an event elicits, characterized by a spectrum from negative to positive [80], [116].

The Backward Digit Span *Task for a 3-Digit* number (BDST3D) was conducted next (step 3), followed by a *Visual Memory Task having 3-Dots* (VMT3D) (step 4) and repeated by a *Backward Digit Span Task for a 4-Digit* number (BDST4D) and a *Visual Memory Task* having 4-Dots (VMT4D). For BDSTs, participants heard a random number, and they had 15 seconds to insert their answers through a projected numerical pad using controllers. Similarly, for the VMT, a random combination of dots in a 3x3 grid was shown for 0.5 seconds, and then participants were asked to memorize and repeat the same pattern through a projected answer sheet in 15 seconds (see Fig. 2). Then, participants were asked to adjust the lighting to their best preference using controllers for the following 30 seconds (step 5). After the lighting adjustment task, the participant experienced five seconds of a black screen, and then the process was repeated for a new condition (step 6). Lastly, after completing ten lighting conditions, the researcher assisted the participant with removing the VR equipment, followed by a post-experiment survey in part 3. 20-30 minutes was assigned to the exit interview time for each participant to discuss the experiment's aims and answer their questions regarding acquired measurements. It should be noted that the results of steps 2 and 5 will be published separately due to their extensive analysis and the need for a more focused discussion.

**Figure 2.** Experiment Sequence. Starting from participant consent, followed by a pre-questionnaire in part 1, a VR session in part 2, a post-questionnaire in part 3, and an exit interview.

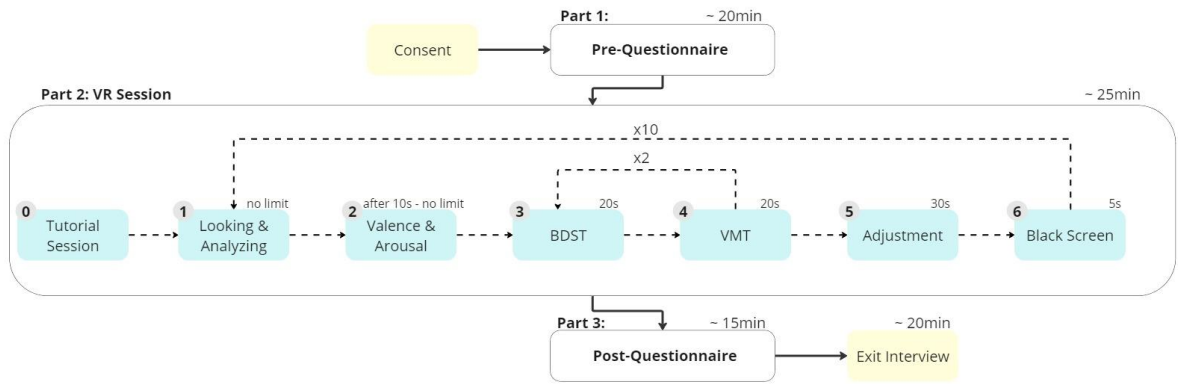

**Figure 3.** The focused view of the participant for BDST and VMT in VR. **A.** Listing to BDST in Cold-Dark (CD) condition, **B.** Answering the BDST in Cold-Bright (CB) condition, **C.** Memorizing VMT4D in Warm-Bright (WB) condition, and **D.** Answering VMT in Neutral-Moderate (NM) condition.

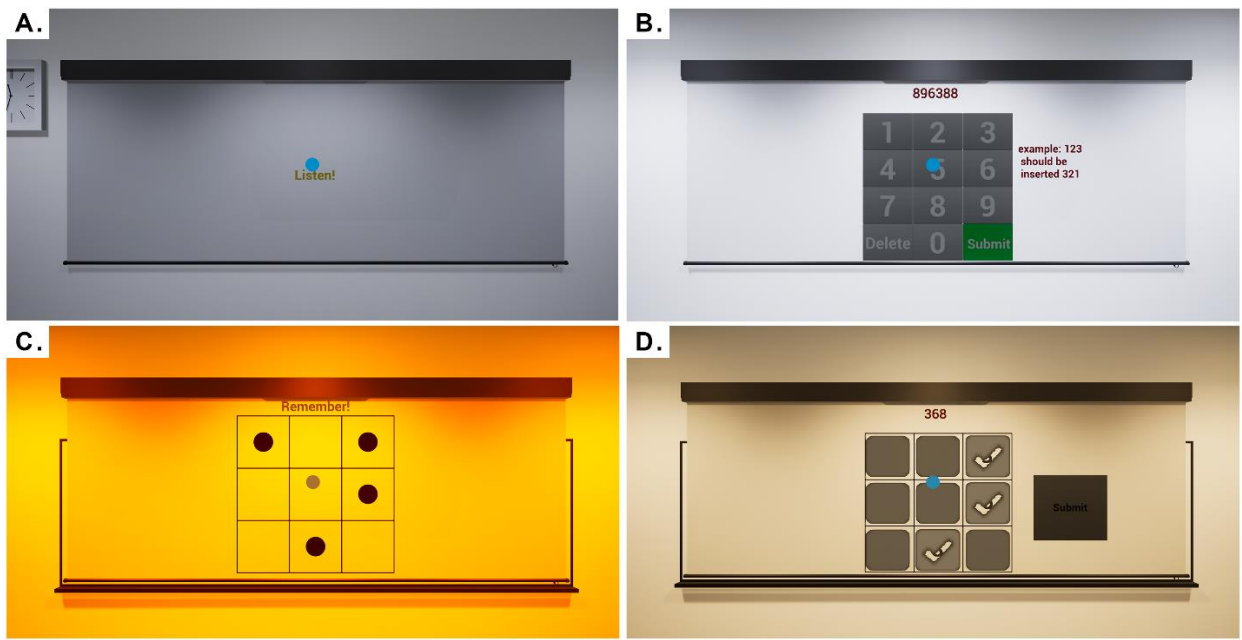

## 2.7 Data Analysis

Due to the nested nature of the data, Mixed Model (MM) analyses were performed to investigate the effects of five given lighting conditions (WD, WB, NM, CD, and CB) on task performance. Data analysis was conducted using the R language, and we used libraries "lme4" and "lmerTest" to fit linear mixed models with fixed effects of LC (Light Condition), AC

(Architectural Context), ET (Experiment Time), as well as LC by AC, LC by ET interaction effects, while controlling for the fixed effect of task difficulty, task order, and the random effect of the participant to predict our dependent measures (BDST score, BDST response time, VMT score, VMT response time), and performed f-tests to examine the interaction effects. Then, the library "emmeans" was used to estimate our dependent measurements at different LC and moderator (if found) levels, followed by pair-wise comparisons (t-tests) with Bonferroni correction. We only controlled for the fixed effects of control variables since we did not expect any interactions between them and lighting conditions (see Table 7 in section 6.3).

## 3. Results

### 3.1 Descriptive Statistics and Potential Confounding Variables

Participants scored 0.88 (SD=0.31) with a mean response time of 10.43s (SD=2.28) for the BDST (Backwards Digit-Span Test) and scored 0.93 (SD=0.22) with a mean response time of 7.29s (SD=2.03) for the VMT (Visual Memory Task). Detailed results for each test in each lighting condition are listed below in Table 3. Moreover, the pre-and post-questionnaire results are presented in section 6.2 (appendices).

**Table 3.** Descriptive Results of Outcome Measures by Group.

|  | CB | CD | NM | WB | WD | Overall |
|---|---|---|---|---|---|---|
| **BDST Score** | 0.88 (0.31) | 0.88 (0.32) | 0.93 (0.26) | 0.89 (0.30) | 0.84 (0.36) | 0.88 (0.31) |
| **BDST RT** | 10.31 (2.29) | 10.48 (2.28) | 10.12 (2.09) | 10.53 (2.27) | 10.69 (2.44) | 10.43 (2.28) |
| **VMT Score** | 0.95 (0.17) | 0.93 (0.22) | 0.90 (0.28) | 0.92 (0.22) | 0.95 (0.18) | 0.93 (0.22) |
| **VMT RT** | 6.97 (1.66) | 7.34 (2.03) | 7.44 (2.45) | 7.51 (2.19) | 7.17 (1.72) | 7.29 (2.03) |

*Note: Mean (SD). The maximum score possible was 1, and the maximum response time was 15 seconds. WD = Warm-Dark condition. CB = Cold and Bright condition, CD = Cold-Dark condition, Neutral and Moderate brightness condition, WB = Warm and Bright condition, CB = Cold and Bright condition, BDST = Backwards Digit-Span Test, VMT = Visual Memory Test, and RT = Response Time.*

### 3.2 Lighting Conditions Affecting Scores in Auditory (RQ1a) and Visual Memory Tasks (RQ1b) Moderated by Architectural Context (RQ1c) and Experiment Time (RQ1d)

Regarding RQ1a, we did not find the main effect of lighting conditions on the BDST score significant. If considering the moderation effect of architectural context and experiment time

(RQ1c and RQ1d), we found the lighting condition by experiment time interaction effect to be significant for BDST, $F(8, 645.99)=2.939$, $p=0.003$. Other interaction effects were not significant. Detailed f-test results can be found in Table 4.

Regarding the moderation effect of experiment time, our result shows that in the **morning**, the NM (Neutral-Moderate) condition has the best test score of 0.95, 95% CI: [0.86, 1.05]; in the **afternoon**, the NM (Neutral-Moderate) condition has the best test score of 0.94, 95% CI: [0.85, 1.02]; in the **evening**, WB (Warm-Bright) condition has the best test score of 0.96, 95% CI: [0.85, 1.06]. However, none of the conditions were significantly better than all the other alternatives at the same experiment time in the post-hoc pairwise comparisons. Lastly, the direction of control variables (i.e., Task Difficulty and Order) is reported in section 6.3.

**Table 4.** F-test Results of Predictors of BDST and VMT Scores

| Predictor | Numerator DF | Denominator DF | F Value | P Value | Partial η2 | ω² |
|---|---|---|---|---|---|---|
| **BDST Score** | | | | | | |
| **Lighting Condition** | 4 | 645.48 | 1.339 | .254 | .008 | .002 |
| **Experiment Time** | 2 | 37.16 | 0.931 | .403 | .048 | <.001 |
| **Architectural Context** | 1 | 649.98 | 0.955 | .329 | .001 | <.001 |
| **LC: ET** | 8 | 645.99 | 32.939 | .003 | .035 | .023 |
| **LC: AC** | 4 | 645.50 | 0.594 | .667 | .004 | <.001 |
| **Task Difficulty** | 1 | 645.31 | 9.715 | .002 | .015 | .013 |
| **Order** | 1 | 647.49 | 11.182 | .001 | .017 | .015 |
| **VMT Score** | | | | | | |
| **Lighting Condition** | 4 | 646.07 | 1.928 | .104 | .012 | .006 |
| **Experiment Time** | 2 | 38.18 | 0.193 | .825 | .010 | <.001 |
| **Architectural Context** | 1 | 650.68 | 2.584 | .108 | .004 | .002 |
| **LC: ET** | 8 | 646.53 | 2.657 | .007 | .032 | .020 |
| **LC: AC** | 4 | 646.09 | 0.673 | .611 | .004 | <.001 |
| **Task Difficulty** | 1 | 645.89 | 0.069 | .793 | <.001 | <.001 |
| **Order** | 1 | 648.14 | 37.134 | <.001 | .054 | .053 |

*Note: Type III Analysis of Variance Table with Satterthwaite's method. LC = Light Condition, ET = Experiment Time, AC = Architectural Context.*

In the case of RQ1b for the effect of lighting conditions on VMT, the effect of lighting conditions on the reaction time was not significant. We found the interaction effect of lighting condition and experiment time significant, $F(8, 646.53)=2.657$, $p=0.007$. Other interaction effects were not significant (See Table 4).

Regarding the VMT scores, our results for the moderation effect of experiment time show that in the **morning**, the WD (Warm-Dark) condition has the best test score of 0.96, 95% CI: [0.90, 1.03]; in the **afternoon**, the NM (Neutral-Moderate) condition has the best test score of 0.96, 95% CI: [0.90,1.03]; in the **evening**, CB (Cold-Bright) condition has the best test score of 0.99, 95% CI: [0.91, 1.06]. However, none of the conditions were significantly better than all the other alternatives at the same experiment time in the follow-up pairwise post-hoc comparisons. Significant and marginally significant differences were observed for the BDST score in the afternoon session (fig 4.A) and for VMT in the morning session (fig 4.B).

**3.3 Lighting Conditions Affecting the Reaction Time in Auditory (RQ2a) and Visual Tasks (RQ2b) Moderated by Architectural Context (RQ2c) and Experiment Time (RQ2d)**

Regarding the impact of lighting conditions on participants' reaction time during the BDST (RQ2a), a marginally significant interaction between lighting conditions and experiment time was found for BDST reaction time, $F(8, 645.48)=1.730$, $p=0.088$ (RQ2c). Other interaction effects were not significant. Detailed f-test results can be found in Table 5.

Our results for the moderation effect of experiment time on BDST reaction time show that in the **morning**, the NM (Neutral-Moderate) condition has the fastest response of 10.29, 95% CI: [9.57, 11.00]; in the **afternoon**, the NM (Neutral-Moderate) condition has the fastest response of 10.10, 95% CI: [9.44, 10.76]; in the **evening**, CB (Cold-Bright) condition has the fastest response of 9.68, 95% CI: [8.89, 10.47]. However, none of the conditions were significantly better than all the other alternatives at the same experiment time in the post-hoc pairwise comparisons.

**Table 5.** F-test Results of Predictors of BDST and VMT Reaction Time

| Predictor | Numerator DF | Denominator DF | F Value | P Value | Partial η2 | ω² |
|---|---|---|---|---|---|---|
| **BDST Reaction Time** | | | | | | |
| Lighting Condition | 4 | 645.39 | 1.661 | .157 | .010 | .004 |
| Experiment Time | 2 | 45.93 | 0.805 | .453 | .034 | <.001 |
| Architectural Context | 1 | 653.30 | 0.324 | .570 | <.001 | <.001 |
| LC: ET | 8 | 645.48 | 1.730 | .088 | .021 | .009 |
| LC: AC | 4 | 645.45 | 0.093 | .985 | <.001 | <.001 |
| Task Difficulty | 1 | 645.05 | 239.741 | <.001 | .271 | .270 |
| Order | 1 | 649.61 | 30.491 | <.001 | .045 | .043 |

|  | VMT Reaction Time | | | | | |
| --- | --- | --- | --- | --- | --- | --- |
| Lighting Condition | 4 | 645.97 | 1.566 | .182 | .010 | .003 |
| Experiment Time | 2 | 43.70 | 1.383 | .262 | .060 | .016 |
| Architectural Context | 1 | 652.92 | 5.580 | .018 | .008 | .007 |
| LC: ET | 8 | 646.13 | 2.265 | .022 | .027 | .015 |
| LC: AC | 4 | 646.01 | 1.486 | .204 | .009 | .003 |
| Task Difficulty | 1 | 645.68 | 3.337 | .068 | .005 | .004 |
| Order | 1 | 649.48 | 69.241 | <.001 | .096 | .095 |

*Note: Type III Analysis of Variance Table with Satterthwaite's method. LC = Light Condition, ET = Experiment Time, and AC = Architectural Context.*

The main effect of lighting conditions on reaction time was not significant for VMT (RQ2b). The interaction effect of lighting condition with the experiment time for VMT was found significant, $F(8, 646.13)=2.265$, $p=0.022$. Other interaction effects were not significant. Detailed f-test results can be found in Table 5.

Regarding the response time for VMT, our results for moderation effect of experiment time show that in the **morning**, the WD (Warm-Dark) condition has the fastest response of 7.20, 95% CI: [6.55, 7.85]; in the **afternoon**, the NM (Neutral-Moderate) condition has the fastest response of 6.93, 95% CI: [6.31, 7.55]; in the **evening**, the CB (Cold-Bright) condition has the fastest response of 6.49, 95% CI: [5.75, 7.23]. It should be noted that none of the conditions was significantly better than all the other alternatives at the same experiment time in the follow-up pairwise post-hoc comparisons. We observed significant and marginally significant differences in RT for BDST in the afternoon and evening session (fig 4.C) and for VMT in the morning and afternoon session (fig 4.D). Regarding the direction of control variables (i.e., Task Difficulty and Order), the results are reported in section 6.3.

**Figure 4**. The differences in cognitive performance during daytime for each dependent measurement with model estimated 95% CIs. **A.** BDST score, **B.** VMT score, **C**. BDST reaction time, and **D**. VMT reaction time. Each horizontal bar shows the lighting conditions with corresponding illuminance and CCT levels and the corresponding score/reaction time. The scores were measured on a scale of 0 (low) to 1 (high), and the reaction time was measured in seconds. Horizontal bars based on pairwise comparisons with Bonferroni correction, '*' = $p<0.05$ and '+' = $p<0.1$.

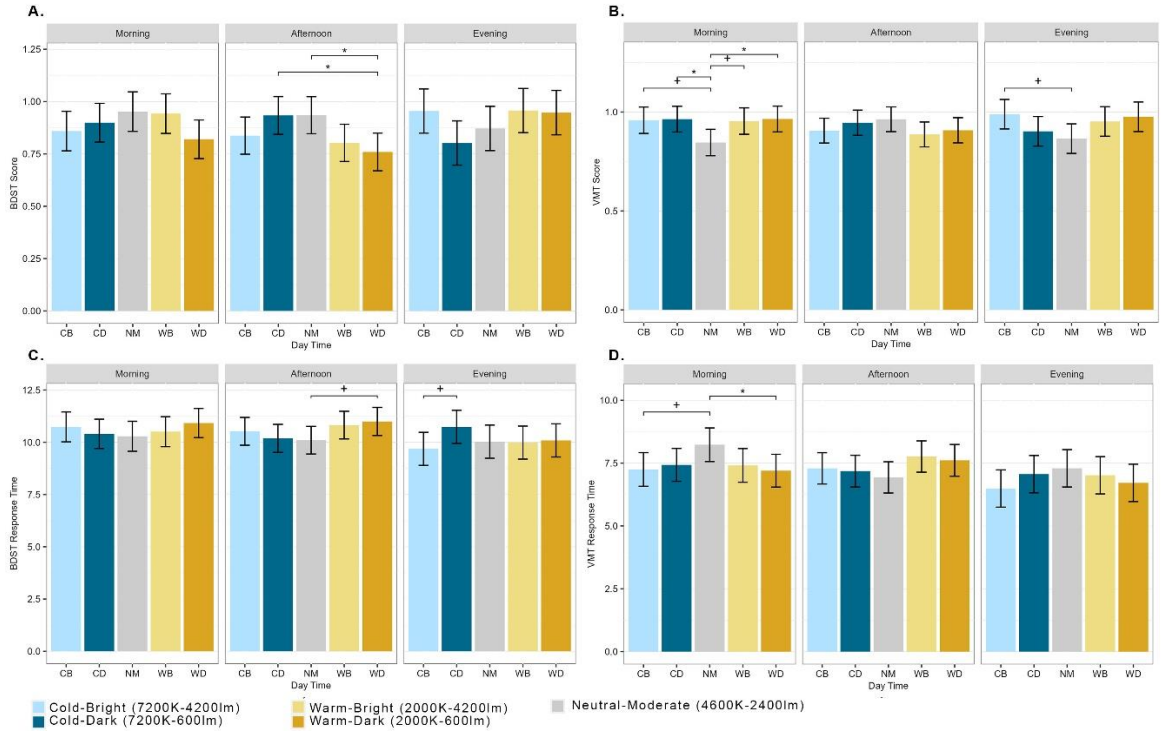

## 3.4 Importance of Intensity and CCT (Exploratory Analysis)

To better understand the importance of illuminance and CCT on task performance, we fitted boosted trees models (xgboost) [118] to estimate our dependent measures with 10-time 10-fold cross-validation and evaluated the importance of the predictors with library "caret" by comparing normalized differences in mean squared errors (MSEs) of models with and without predictor averaged over all trees. The results suggested CCT to be a more important predictor except for the BDST score. However, similar to effect sizes, other predictors were found to have higher importance (table 6).

**Table 6.** Importance of Predictors of BDST and VMT Score and Reaction Time

| Predictor Importance | BDST Score | BDST RT | VMT Score | VMT RT |
|---|---|---|---|---|
| Daytime: Afternoon* | 0.130 | <.001 | 0.007 | <.001 |
| Daytime: Evening* | <.001 | 0.065 | <.001 | 0.090 |
| Difficulty level | 0.205 | 0.755 | <.001 | 0.023 |
| CCT | 0.024 | 0.015 | 0.031 | 0.013 |
| Illuminance | 0.065 | 0.006 | 0.014 | 0.005 |
| Architectural Context | 0.028 | 0.001 | 0.007 | 0.018 |
| Order | 0.548 | 0.157 | 0.941 | 0.851 |

*Note: Daytime with three levels was dummy-coded with reference-level coding. Daytime: Afternoon = 0, Daytime Evening = 0 when Daytime = "morning"; Daytime: Afternoon = 1, Daytime Evening = 0 when Daytime = "afternoon"; Daytime: Afternoon = 0, Daytime Evening = 1 when Daytime = "evening".

# 4. Discussion

The main goal of this study was to evaluate the effect of different lighting conditions (LCs) on users' cognitive performance. The users' cognitive performance was evaluated by conducting auditory (BDST) and visual memory (VMT) tests at two low difficulty levels. The analysis of the test was based on both participants' scores and the corresponding reaction time (RT) for each task. The second goal of this study was to examine the moderation effect of architectural context and time of the experiment.

The discussion section has been organized in alignment with the order of the research questions and the conducted experiment. To answer the research questions, we designed five different LCs with different levels of light intensities and CCTs, while the LCs were identically tested in two different architectural contexts (ACs). Participants experienced these ten LCs randomly and answered four different tests with two different difficulty levels. The conducted within-subject experiment provided a cluster of measurements from each participant, with daytime as a between-subject factor. Accordingly, for each outcome variable in RQ1 and RQ2, it was investigated whether the effect of LCs (i.e., illuminance and CCT) was moderated by architectural context and daytime while controlling for the order of presented LCs and task difficulty level.

## 4.1 Contribution to Lighting Research

Overall, the findings related to RQ1 indicate that different lighting conditions did not significantly affect auditory and visual memory task scores. Participants performed well across all lighting conditions, suggesting no direct impact of lighting manipulations on task performance. However, we observed an interaction effect with experiment time, aligning with previous research [112], [113], [114]. The role of daytime in light effects on cognitive performance remains inconclusive in the literature, with mixed findings [20], [27], [30], [47], [115], [116] [23], [117]. Our study suggests that the influence of lighting conditions on audio and visual task performance varies with the moderation effect of daytime, with significant differences observed during afternoon sessions for auditory tasks (BDSTs) and morning sessions for visual memory tasks (VMTs).

Additionally, the main effect of architectural context was significant for visual memory task reaction time, with participants responding faster in the classroom compared to the meeting room. Lighting condition order and task difficulty were controlled for all dependent variables, revealing a significant learning effect over time. To be precise, increasing task difficulty by adding an extra digit to BDST decreased scores and increased reaction time, while task difficulty did not significantly affect visual memory task scores.

Given our execution of low-difficulty BDSTs, the outcomes in virtual reality align with those observed in a physical lab environment. This consistency is reflected in two studies by Huiberst et al. [30], [47], where the comparison of light intensity did not yield a significant difference in BDST scores. [30], [44]As the data were collected during the period of two weeks in December 2022, our BDST results could also be compared against easy trials of those in Heuibers et al.'s seasonal study [17]. Using VR, we found a partially similar trend in the case of the morning while comparing the illuminance level in which participants scored higher in Neutral-Moderate (4600 K-400lx) rather than Cold-Dark (7200 K-100 lx) and Warm-Dark (2000 K-100 lx). However, we could not compare the afternoon session with Heuibers et al.'s results [17] as the afternoon experiment time sessions did not match the ones from previous research. Moreover, similar to their study, we also confirmed that the effect of LCs on easy BDST scores is not significant during the different daytime.

Another finding for BDSTs was the significant effect of CCT in the afternoon sessions. Regardless of illuminance, the warm light has resulted in significantly lower BDST scores in the afternoon session while having the Neutral-Moderate (4600 K-400 lx) condition. Regardless of the time of the day, the impact of CCT on BDST scores is partially in line with those found in Yang and Jeon's study [13]. We can confirm that in an afternoon session, warm light (2000 K) led to a lower BDST score in comparison with natural (4600 K) and cold (7200 K) light. The results of the BDST are partially in line with those findings for the working memory task in Ru et al.'s study [23] in which higher brightness, 400 lx vs. 100 lx, was used in our study (instead of 1000 lx vs.

100 lx used in the referenced study), would result in higher cognitive performances. However, we did not find significant differences between 700 lx (WB and CB) and 400 lx (NM).

Regarding the visual task scores, our result showed that the bright light (i.e., WB and CB, 700 lx) improves VMT scores compared to moderate lighting (i.e., NM, 400 lx) in the morning and evening sessions and not in the afternoon sessions. A similar trend was found for dark conditions (100 lx) compared to moderate light intensity (400 lx). We assume this was caused mainly by having a higher contrast ratio in the VMT. Regarding the CCT, the results for comparing ~4000K vs. ~7000K are also, to a certain extent, in line with those findings in Zeng et al. [41], showing the score improvement for higher CCT. Regarding warm light, our results are in line with those found by Sahin et al. [27] for the morning and evening time, where lower CCT increased the visual memory performance of the visual task.

Regarding the results for the lighting affecting the reaction time for BDST and VMT (RQ2), we found a similar trend as in RQ1. The reaction time for BDSTs was not affected significantly in different lighting conditions (only marginally significant for reaction time for VMTs). Moreover, our results revealed that higher scores were mainly linked with faster reaction time for each lighting condition during each experiment time. Without considering the daytime effect, our VR results for auditory BDSTs align with those found by [47], [118] for any n-back test.

In the morning session, reaction time was significantly faster for Cold-Bright and Warm-Dark than Neutral-Moderate conditions, indicating quicker responses during visual memory tasks. In this regard, our data is in line with those in Ru et al.'s study [23] of Go/No-go tasks. If we consider the effect of daytime, then our results for the VMT are partially in line with those for PVT in Smolders et al.'s study [119]. We could compare our result with findings in previous studies for the morning, in which participants respond faster overall in bright conditions, while comparisons for evening and afternoon are not possible due to the time slot differences.

Lastly, in our analysis, boosted trees models indicated that CCT emerged as a more crucial predictor than illuminance for various cognitive performance measures, including test scores and reaction times. This pattern held true for all aspects except for the BDST score, emphasizing the potentially underappreciated role of CCT in influencing cognitive performance in working and educational environments.

### 4.2 Implications for Practice

The research results show a statistically significant dependence for the interaction effect between the time of the day when the experiment took place, cognitive test performance scores, and reaction time. The interaction effects on task performance found in this study had medium-small effect sizes. Practically, this suggested that while such effects exist, the change in cognitive performance caused by the interaction effects was not that large. The relatively small effect size, combined with the conventional 0.80 power requirement, could also possibly explain the inconsistent findings in previous studies. These results indicate the importance of further research into the role of lighting in educational spaces, primarily focusing on daytime contribution and task type during exposure to specific light conditions.

Additionally, evidence-based design methods can become more affordable and accessible as pre-construction testing methods by carefully analyzing user responses to lighting factors in virtual reality. This will eventually produce enormous comparative data sets across a variety of building designs and participants. Another notable implication of our method is that it could tailor the lighting setting for specific space use during the daytime using the knowledge of user cognitive performance. Such an approach can cluster the results with a precise timestamp for both auditory and visual memory tasks and be integrated with other data, such as physiological measurements [61]. Finally, combining VR with lighting provides exciting possibilities for lighting research with fully customizable environments for various user interactions.

### 4.3 Limitations and Future Works

While this study was dedicated to developing immersive lighting simulations in UE4, it is essential to recognize that the measurements utilized in the study have yet to be validated against real-world measurement methods. It is important to note that these measurements should not be directly applied in real environments. However, this presents an opportunity for future validation and refinement of our simulations to align them with real-world applications. Nonetheless, given the aim of our experiment to compare the impact of illuminance and CCT on cognitive performance and confirm the moderation effect of daytime using VR, the potential limitations of our simulation could be a minor concern. Therefore, further research, such as this study, is required to validate and discuss these findings in relation to real-world environments.

Similar to many relevant studies conducted in the field, we utilized a participant pool consisting mainly of university students and were hence limited by convenience sampling. As a result, the findings should be interpreted with caution and are possibly limited to the sampled sub-population. Ideally, a follow-up study with systematical sampling on the world population, possibly taking the variability of strata into consideration, would give us an unbiased estimation of the effects of lighting conditions. While we do not expect education level/field of study to have any major influence on task performance, the age distribution (all participants were young adults) and possibly the proportion of participants wearing eyeglasses could be different from the general population and may lead to difference effects of lighting condition on task performance.

When constraints are appropriately addressed, VR can be used as a viable exploration tool for study and design in the lighting sector. While VR offers greater control over lighting than real-world settings, it can also have a limit on the number of participants who can be tested at once. This can make it challenging to draw broad conclusions about how lighting affects large numbers of people. Another limitation for generalizing our results was the constraint in the geographical location during the data collection [120] which can be repeated by conducting our experiment in different countries. Moreover, we performed the experiment using relatively low-difficulty tasks, and we will continue our experiment by having more difficult trials for BDST and VMT under similar LCs. Lastly, the role of architectural context will also be investigated in our future studies, which will involve more difficult trials in VR.

## 5. Conclusions

This study examined the effects of lighting conditions, architectural context, and time of day on cognitive performance in human participants. The findings of this study indicate that the influence of different lighting conditions on audio and visual task performance varies. Notably, significant differences in audio task performance (BDSTs) were primarily observed during afternoon sessions, whereas significant differences in visual memory task performance (VMTs) were more pronounced during morning sessions. This confirms that the effects of lighting conditions on task performance are significantly moderated by the time of the experiment. Moreover, our results suggested that CCT is a more important predictor compared with illuminance for all the tests (i.e., VMT scores and reaction times) except for the BDST score.

Furthermore, the study revealed a significant main effect of architectural context on reaction time for visual memory tasks. Specifically, participants exhibited faster reaction times in the classroom than in the meeting room. However, no significant interaction between lighting conditions and architectural context was observed, indicating that the impact of lighting conditions on task performance did not vary significantly between the two architectural contexts.

These findings provide valuable insights into the correlation between lighting conditions, task performance, and experiment time. The differential effects of lighting conditions on audio and visual tasks highlight the importance of considering the specific cognitive demands of different tasks when designing lighting environments. Furthermore, the noteworthy main effect of architectural context implies the necessity for additional research to confirm that the environment's physical attributes can influence task performance in conjunction with lighting conditions.

## 6. Appendices

### 6.1 The VR scene

In a dark room, we utilized a Minolta LS 110 luminance meter directed towards the device, similar to Rockcastle et al.'s study [67] (as shown in Figure 5). The maximum luminance recorded with a white image (RGB 255, 255, 255) was 95 cd/m2, while the minimum, under control conditions, was 0.23 cd/m2, observed with a black image (RGB 0, 0, 0). We recorded the luminance every 45 degrees in all lighting conditions and both architectural contexts.

**Figure 5**. The amount of light emitted from the Meta Quest 2 VR headset. **A.** recorded values (cd/m2) for different lighting conditions in meeting room **B.** recorded values (cd/m2) for different lighting conditions in classroom **C.** The luminance meter mounted on the right lens of Meta Quest 2 for taking luminance measurements.

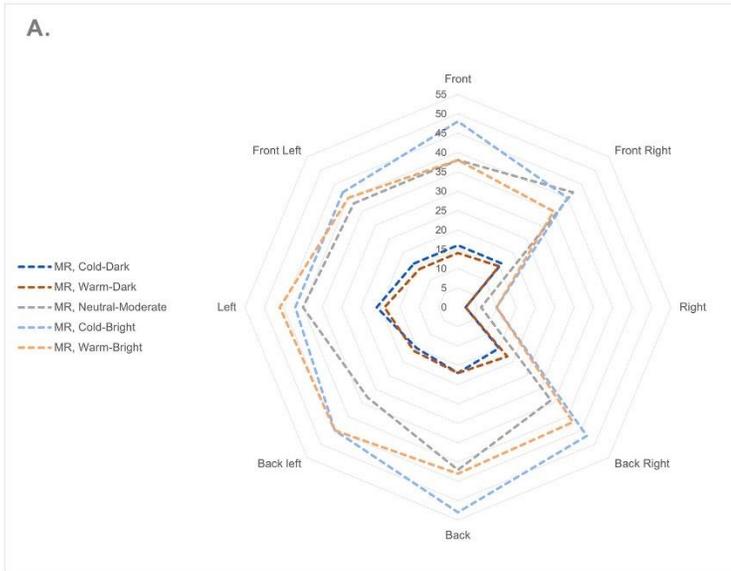

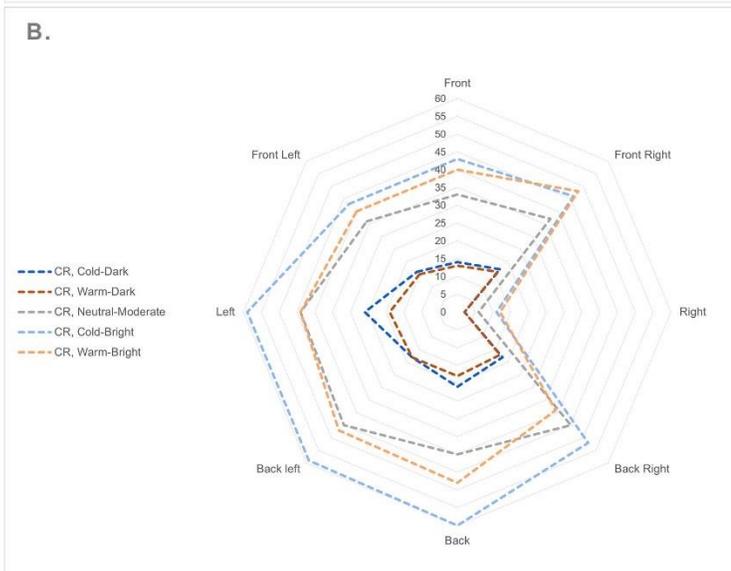

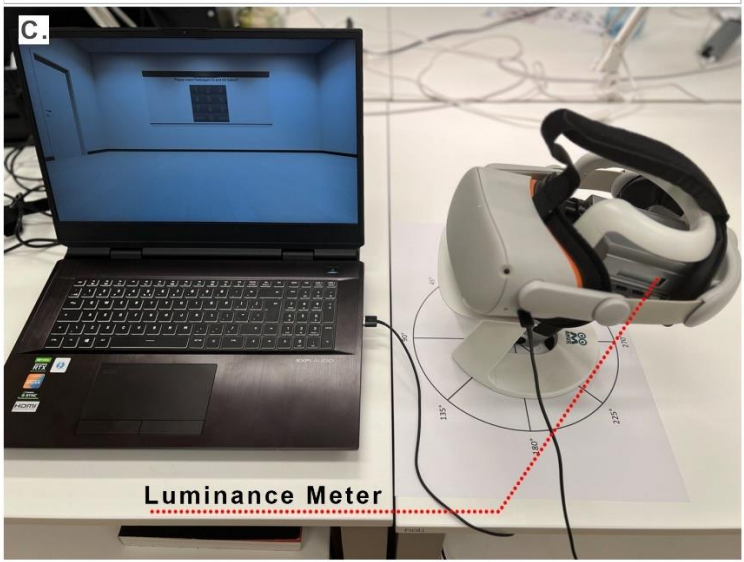

## 6.2 Pre- and Post-questionnaires Results

The average score on the Morningness–Eveningness Questionnaire (MEQ) was 49.57 (SD=5.69), with most of the participants scoring as intermediate (82%). An intermediate score on the MEQ indicates a moderate preference for neither morningness nor eveningness, suggesting a flexible and adaptable chronotype. A smaller number of participants scored as moderate evening (9%) or moderate morning (9%).

Average scores on the IPQ (1–7 scale) were 5.74 (SD=1.13) for the overall sense of presence in the virtual environment, 4.38 (SD=1.18) for the sense of spatial presence, 4.67 (SD=1.82) for the sense of involvement, and 3.62 (SD=1.21) for experienced realism. These above-average results may have positive implications for the effectiveness of our VR environment. The average total score for the SSQ was 26.39 out of 235.62, which shows a small amount of cybersickness caused by our VR environment for the participants. This includes 13.62 for nausea, 32.21 for oculomotor disturbance, and 13.92 for disorientation. The results for the NASA Task Load Index (1–10 scale) indicated a low average score of 4.33 (SD=2.19), showing the medium-high mental workload demand. Reporting on the MEC-SPQ (1-5 scale), the average score for the spatial situation (SSM) was 4.59 (SD= 0.63), showing a relatively high perception of the environment. This was 3.69 (SD= 0.95) for self-location (SPSL), showing the average presence feeling for the participant in the VE, and the average score of 3.26 (SD= 1.27) for the possible action (SPPA) and involvement within the environment.

## 6.3 Controlled Variables Directions

**Table 7.** The direction of control variables (i.e., lighting order and task difficulty)

| Task | Control variable | Coefficient (95% CI) | t | df | P value |
|---|---|---|---|---|---|
| BDST score | Task Difficulty | -0.07 (-0.11, -0.03) | -3.12 | 647.49 | .002 |
|  | Order | 0.01 (0.01, 0.02) | 3.34 | 645.31 | <.001 |
| VMT Score | Task Difficulty | 0.00 (-0.03, 0.03) | 0.26 | 645.89 | .793 |
|  | Order | 0.02 (0.01, 0.02) | 6.09 | 648.14 | <.001 |
| BDST RT | Task Difficulty | 2.07 (1.81, 2.32) | 15.48 | 645.05 | <.001 |
|  | Order | -0.13 (-0.18, -0.09) | -5.52 | 649.61 | <.001 |
| VMT RT | Task Difficulty | 0.24 (-0.01, 0.50) | 1.83 | 645.68 | .068 |
| Order | -0.20 (-0.24, -0.15) | -8.32 | 649.48 | <.001 | |